\documentclass{elsart4}

\usepackage{graphicx}
\usepackage{wrapfig}

\usepackage{amssymb}

\usepackage[english,francais]{babel}

\newtheorem{e-proposition}[theorem]{Proposition}

\newtheorem{e-definition}[theorem]{Definition\rm}

\renewcommand{\theequation}{\arabic{equation}}
\def\og{\leavevmode\raise.3ex\hbox{$\scriptscriptstyle\langle\!\langle$~}}
\def\fg{\leavevmode\raise.3ex\hbox{~$\!\scriptscriptstyle\,\rangle\!\rangle$}}

\begin{document}

\centerline{GRB studies in the SVOM era}
\begin{frontmatter}


\selectlanguage{english}
\title{The {\it Fermi} view of Gamma-Ray Bursts}


\selectlanguage{english}
\author[fp]{F. Piron} and
\ead{piron@in2p3.fr}
\author[vc]{V. Connaughton}
\ead{valerie@nasa.gov}

\address[fp]{Laboratoire Univers et Particules de Montpellier, Université Montpellier 2, CNRS/IN2P3, place Eugène Bataillon, 34095 Montpellier cedex 5, France}
\address[vc]{Center for Space Plasma and Aeronomic Research (CSPAR), University of Alabama in Huntsville, AL 35899, USA}


\medskip

\begin{abstract}
Since its successful launch in June 2008, the {\it Fermi} Gamma-ray Space Telescope has made important breakthroughs in the understanding of the Gamma-Ray Burst (GRB) phenomemon.
The combination of the GBM and the LAT instruments onboard the {\it Fermi} observatory has provided a wealth of information from its observations of GRBs over seven decades in energy.
We present brief descriptions of the {\it Fermi} instruments and their capabilities for GRB science,
and report highlights from {\it  Fermi} observations of high-energy prompt and extended GRB emission.
The main physical implications of these results are discussed, along with open questions regarding GRB modelling.
We emphasize future synergies with ground-based \v Cerenkov telescopes at the time of the SVOM mission.


\vskip 0.5\baselineskip
\selectlanguage{francais}
\vskip 0.5\baselineskip
\noindent
{\bf Observations des sursauts gamma avec Fermi (r\'esum\'e)}
\vskip 0.5\baselineskip
Depuis sa mise en orbite en juin 2008, le t\'elescope spatial {\it Fermi} a permis des avanc\'ees remarquables dans la compr\'ehension des sursauts gamma.
La moisson de r\'esultats obtenus par {\it Fermi} a \'et\'e rendue possible par la combinaison des instruments \`a bord de l'observatoire, le GBM et le LAT, couvrant un domaine spectral s'\'etendant sur plus de sept ordres de grandeur en \'energie.
Nous r\'esumons les caract\'eristiques des deux instruments et leurs capacit\'es pour la d\'etection et l'\'etude des sursauts gamma, passons en revue les r\'esultats les plus marquants, et pr\'esentons leurs implications physiques imm\'ediates.
Apr\`es un rapide examen des questions soulev\'ees par ces observations et des enjeux th\'eoriques futurs, nous discutons les synergies observationnelles
avec les t\'elescopes qui seront op\'erationnels aux tr\`es hautes \'energies \`a l'\`ere de la mission SVOM.
%

 \vskip 0.5\baselineskip
\noindent{\small{\it Mots-cl\'es~:} sursauts gamma; {\it Fermi}; facteur de Lorentz d'ensemble; fond diffus cosmique; invariance de Lorentz; t\'elescopes \v Cerenkov
}}
\end{abstract}
\end{frontmatter}


\selectlanguage{english}

\section{Introduction}     
Before the era of the {\it Fermi} Gamma-ray Space Telescope, high-energy emission from Gamma-Ray Bursts (GRBs) was observed with 
the Energetic Gamma-Ray Experiment Telescope (EGRET, covering the energy range from 30~MeV to~30~GeV) onboard the Compton Gamma-Ray Observatory (CGRO; 1991--2000) and, more recently,
by the GRID instrument onboard Astro-rivelatore Gamma a Immagini LEggero (AGILE)~\cite{giuliani08}.
Despite poor photon statistics owing to the effective area and deadtime limitations of EGRET, substantial emission above 100~MeV was detected in a few distinct cases. 
Together, these cases illustrate the diversity in GRB spectral and temporal properties at high energies, and the advances provided by {\it Fermi} were eagerly anticipated.  
The high-energy emission from GRB~930131 was consistent with an extrapolation from the typical keV--MeV spectrum~\cite{sommer94} observed with EGRET's on-board partner, the Burst And Transient Source
Experiment (BATSE).
In the case of GRB~941017, evidence was found for an additional, hard spectral component 
extending up to $\sim$200~MeV and lasting significantly longer ($\sim$200~s) than the low-energy spectral component seen with BATSE~\cite{gonzalez03}.
GRB~940217, for which delayed high-energy emission was detected up to $\sim$90 minutes after the BATSE trigger time, including an 18~GeV photon detected after $\sim$75 minutes~\cite{hurley94}, was the most compelling case that the high-energy emission of GRBs provided a new window to GRB physics.\\
\begin{wrapfigure}{l}{0cm}
\centering
\includegraphics[width=.4\linewidth]{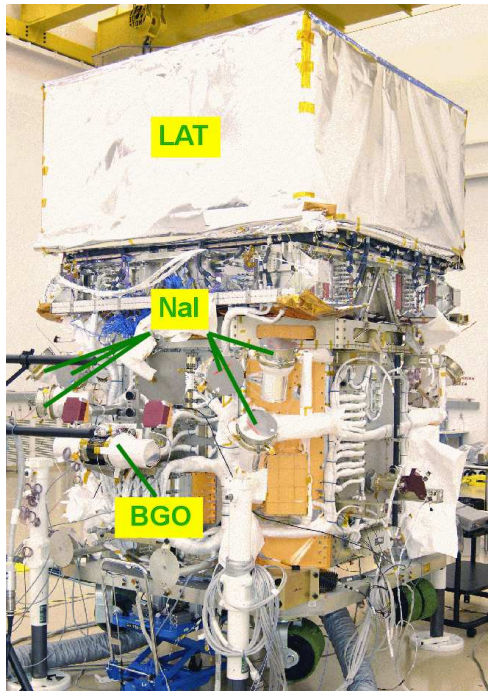}
\caption{
The {\it Fermi} observatory.
}
\label{Fig:Observatory}
\end{wrapfigure}

Following the steps of its predecessors, the {\it Fermi} observatory was placed into orbit on June $11^{th}$ 2008 for a 5 to 10 year mission, and provides an unprecedented energy coverage and sensitivity for advancing our knowledge of GRB properties at high energies.
It is composed of two instruments (see Figure~\ref{Fig:Observatory}), the Gamma-ray Burst Monitor (GBM; \cite{meegan09}) and the Large Area Telescope (LAT; \cite{atwood09}), which together cover more than 7 decades in energy.
The GBM is comprised of 14 scintillation detectors which monitor the entire sky that is not occulted by the Earth. 
Triggering and localization are performed using twelve Sodium Iodide (NaI) detectors placed in 4 groups of 3 around the spacecraft, the different orientations of each detector allowing reconstruction
of the source position to within a few degrees accuracy.
The GBM spectroscopy makes use of both the NaI detectors between 8~keV and 1~MeV and two Bismuth Germanate (BGO) scintillators which are sensitive to photons of energies between 150~keV and 40~MeV.
As a result, the GBM can measure spectra with high time resolution over nearly 5 decades in energy, and provides a bridge from
the low energies (below $\sim$1~MeV), where most of the GRB emission takes place, to the less well-explored territory accessible to the LAT.

The LAT is a pair production telescope sensitive to gamma rays in the energy range from 20~MeV to more than 300~GeV.
The telescope consists of an array of 4$\times$4 identical towers, each made by a tracker of silicon strip planes with slabs of tungsten converter, followed by a module of a hodoscopic Cesium Iodide (CsI) calorimeter.
This array is covered by a segmented anti-coincidence detector which is designed to efficiently identify and reject charged
particle background events. 
The LAT broad energy range, large effective area ($\sim$8000~cm$^2$ at peak), low deadtime per event ($\sim$27~$\mu$s), wide field-of-view ($\sim$2.4~sr at 1~GeV) and good angular resolution ($\sim$0.15$^\circ$ at 10~GeV) are vastly improved in comparison with those of EGRET.
They provide more GRB detections, more photons detected from each burst, and precise GRB locations ($\lesssim$1$^\circ$).  The LAT can trigger on GRBs, either autonomously or, with a lower threshold,
 via cross-instrument communication when the GBM is triggered.\\

After two years of operations, the GBM has detected $\sim$500 triggered GRBs~\cite{paciesas10}, with $\sim$50\% occurring in the LAT field-of-view.
In this sample, the LAT has significantly detected $\sim$20 GRBs\footnote{See the complete LAT GRB table: 
http://fermi.gsfc.nasa.gov/\-ssc/\-resources/\-observations/\-grbs/\-grb\_table}, of which one was detected onboard and the rest were recovered in ground analysis.
The ground processing seeks LAT counterparts to known GRBs, which previously triggered the GBM and/or other instruments.
It also performs a blind search for bursts not detected by other instruments, but no new GRB has been found by this procedure so far.
Onboard triggers are especially desirable given the latency of up to 12 hours for discovery and localization on the ground versus a few seconds for an
onboard localization, and efforts to increase this onboard trigger rate should result in a future yearly rate of 3 to 5~\cite{gcn10777}.
Owing to the detection of extended emission by EGRET from GRB~940217, and the interest in GRB afterglow emission during the {\it Swift} era, {\it Fermi} was designed with the capability to repoint in the direction of a bright GRB and keep
its position near the centre of the field-of-view of the LAT for, nominally, five hours, subject to Earth-limb constraints.
This repointing occurs autonomously in response to requests from either instrument,
with adjustable brightness thresholds, and has resulted in 45 extended GRB observations since October 2008 when the capability was enabled.  
Both GBM and LAT triggers and localizations are communicated to other satellites and ground observers
via the GRB Coordinates Network~\cite{gcn}, and all {\it Fermi} data are public through the {\it Fermi} Science Support Center~\cite{fssc}.\\

In Section~\ref{Sec:Observations} we summarize the main results obtained with the GBM and the LAT, and present the properties of GRBs as revealed by the two instruments.
The physical implications of these observations are addressed in Section~\ref{Sec:Implications}.
In Section~\ref{Sec:Perspectives}, we discuss several open questions and topics of interest for the near future.
We also present some perspectives for the forthcoming years and until the advance of the SVOM mission, including the expectations from ground-based \v Cerenkov experiments operating at $\sim$100~GeV--TeV energies.

\section{Observations}
\label{Sec:Observations}
\begin{figure}[t!]
\includegraphics[width=.5\linewidth]{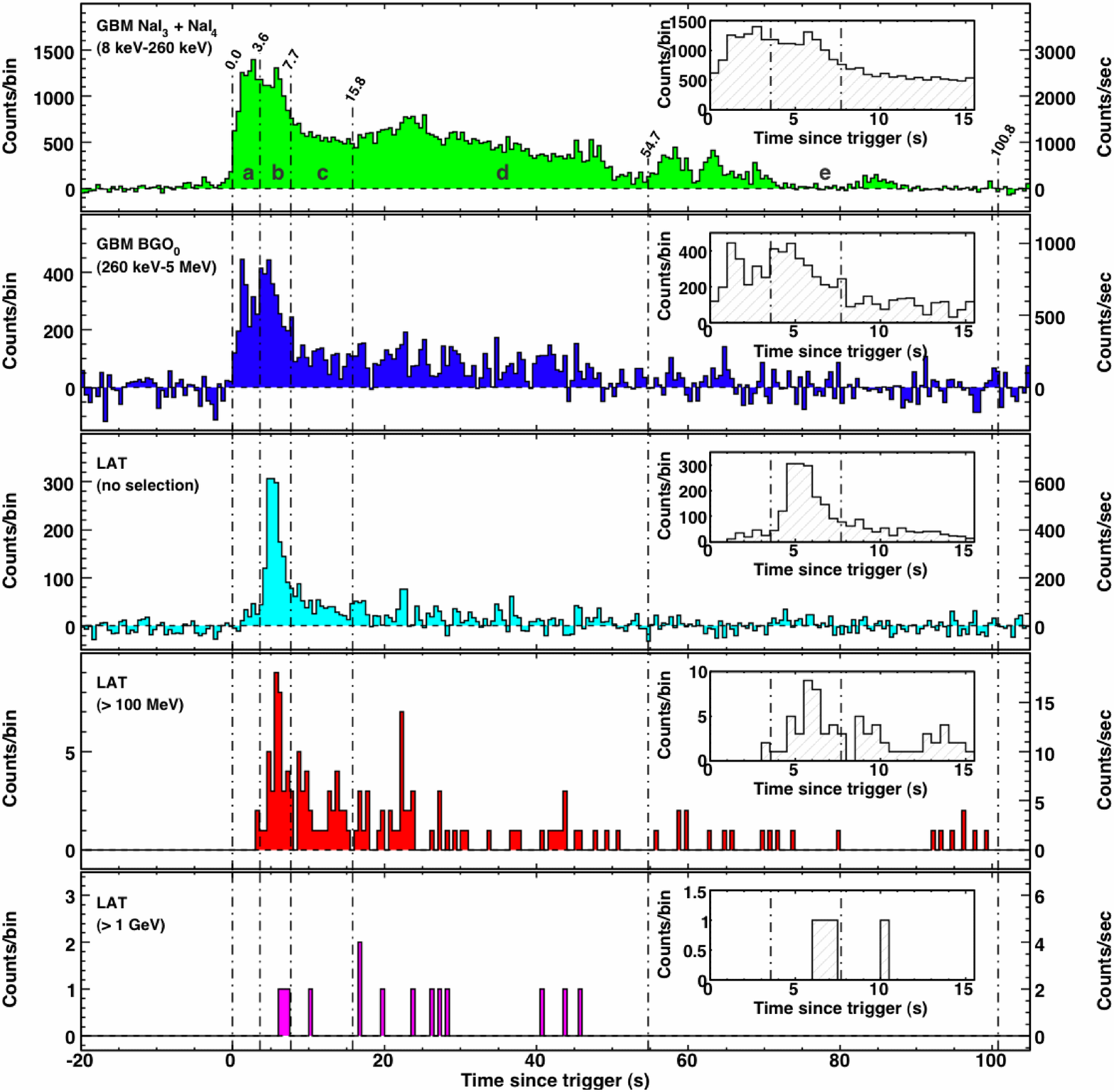}
\includegraphics[width=.5\linewidth]{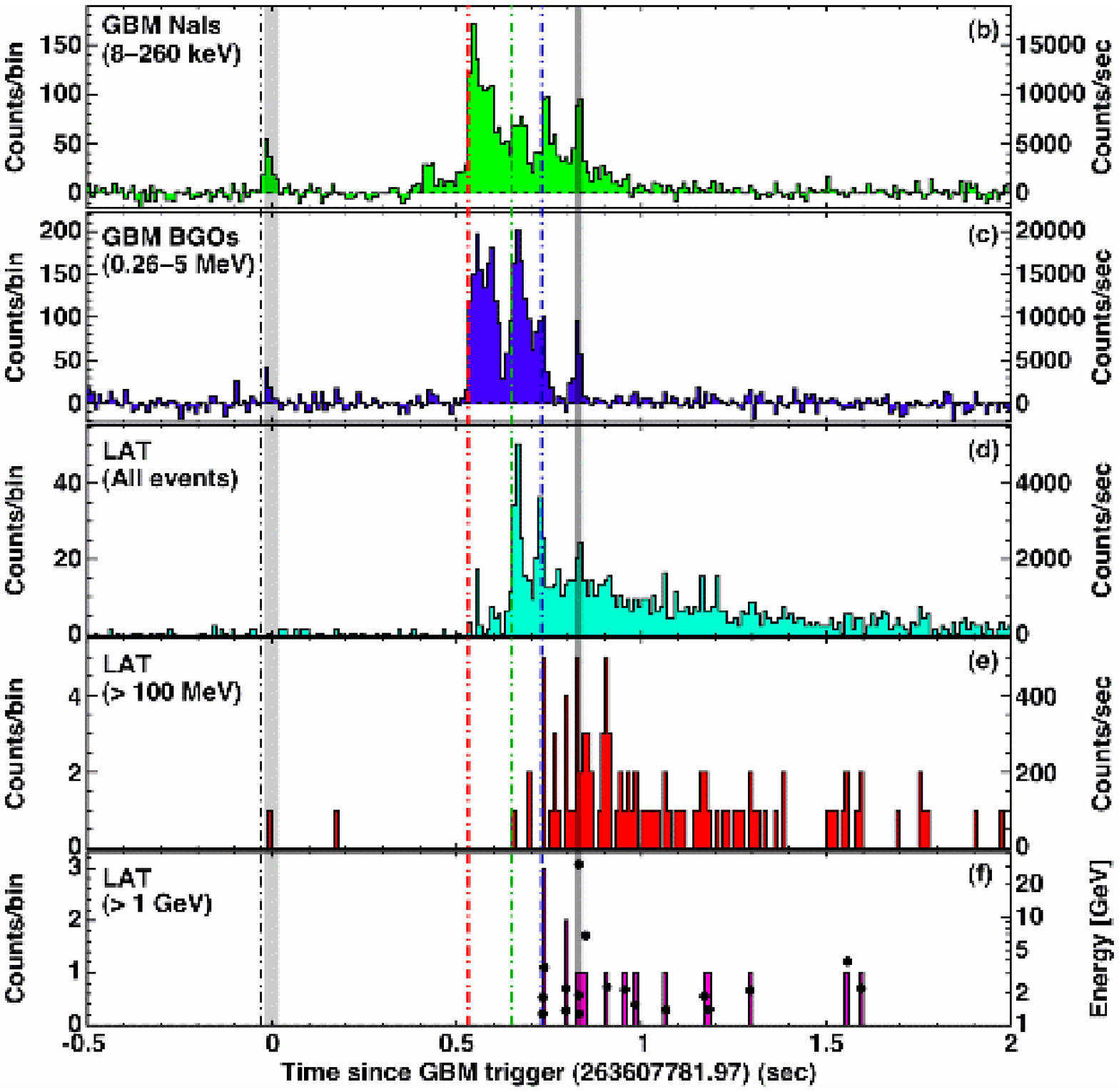}
\caption{
{\it (Left)} Light curves for GRB~080916C observed with the GBM and the LAT, from lowest to highest energies~\cite{080916c}.
The energy ranges for the top two panels are chosen to avoid overlap.
The top three panels represent the background-subtracted light curves for the NaI, the BGO and the LAT.
The inset panels give a view of the first 15 s from the trigger time. In all cases, the bin width is 0.5~s; the per-second counting rate is reported on the right for convenience.
{\it (Right)} Light curves for GRB~090510~\cite{090510_prompt}, with a time binning of 0.01~s.
}
\label{Fig:LC}
\end{figure}
In two years of observation by {\it Fermi}, the LAT has detected $\sim$20 GRBs with high significance.  
The proportion of short to long bursts detected above 100~MeV is similar to 
that below 1~MeV, about 15-20 \%, though with only two short high-energy bursts, statistics are poor.
In the context of short and long bursts having different progenitors, distances, fluences, and spectral parameters, this consistency may be surprising.  
Moreover,  the steepness of the high-energy spectral index in short bursts detected with the GBM~\cite{guiriec10} would {\it a priori} make a high-energy detection unlikely.  

Four of the LAT-detected bursts, including one of the two short bursts,
are very bright, with hundreds of high-energy photons seen by the LAT, and have led to surprising conclusions regarding
the general properties of the high-energy emission of GRBs.  The remaining bursts, with 20 or fewer photons, 
either confirm these conclusions with less conviction, or at least do not contradict the general observations
we make here.  
The bursts detected by the LAT span a wide range of redshifts, measured by
optical telescopes chasing their afterglow radiation, from z=0.9 to z=4.3, implying for the most
distant of the LAT bursts, GRB~080916C, the largest apparent isotropic energy release ever, $E_\mathrm{iso}$$\simeq$8.8$\times 10^{54}$ ergs
assuming a standard cosmology~\cite{080916c}.\\

Figure~\ref{Fig:LC} shows the GBM and LAT light curves for two of the bright LAT-detected bursts, with the count rates
as a function of time for the lowest to highest energies displayed from top to bottom.  Although the time-scales
are different for these long (GRB~080916C, left) and short (GRB~090510, right) bursts, the pattern is the same: the
first peak is missing in the LAT light curve above 100~MeV, though the first and second peaks are of similar brightness
at low energies in GBM. This is a new and unexpected result from {\it Fermi}, and is clearly seen in all four
bright LAT-detected bursts. It is one of the dimmer bursts, GRB~090217, however, that provides the clearest
evidence that this missing first peak in the LAT may be a result of soft-to-hard spectral evolution rather than a cut-off
in the spectrum below 100~MeV during the initial peak of a GRB~\cite{090217}.
In GRB~090217, the first and second peaks seen by GBM are spectrally similar, and 
extend into the LAT range without evidence for cut-offs, though the brightest peak in the LAT appears
later than the brightest peak in GBM, consistent with the initial soft-to-hard evolution observed in the brighter bursts.
\begin{figure}[t!]
\includegraphics[width=.47\linewidth]{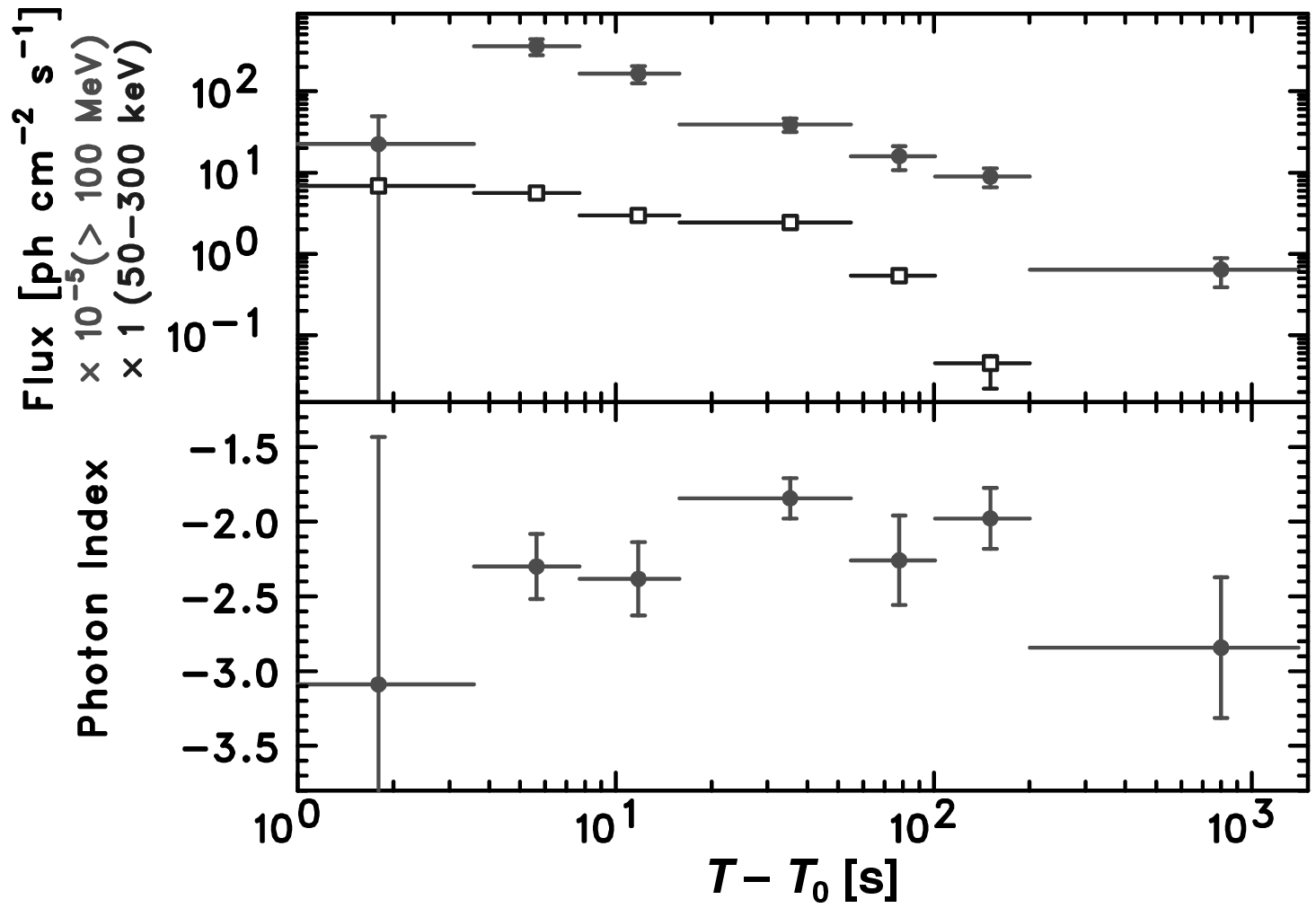}
\includegraphics[width=.53\linewidth]{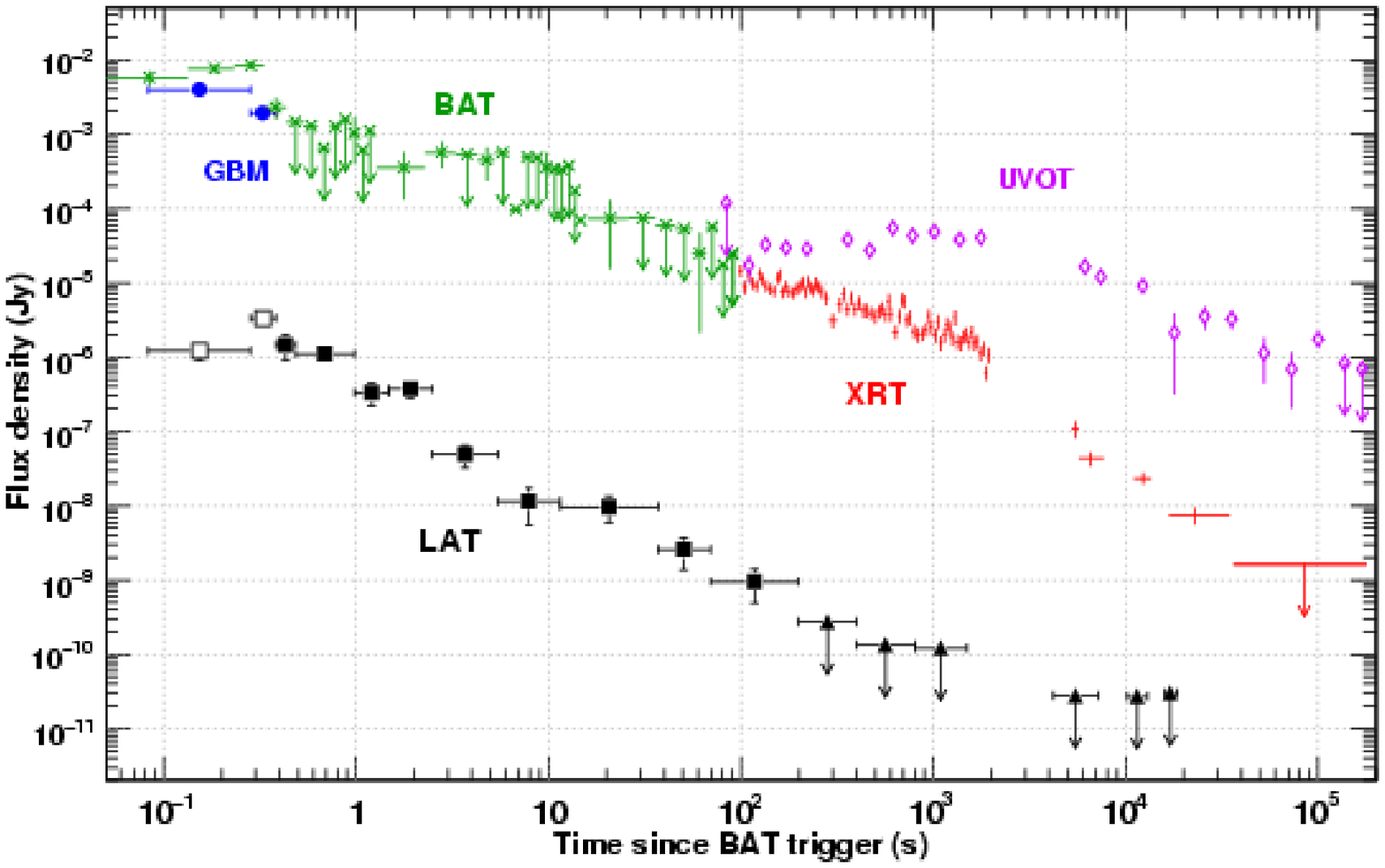}
\caption{
{\it (Left)} GRB~080916C fluxes (top panel) for the energy range 50--300~keV (open squares) and above 100~MeV (filled symbols), and power-law index as functions of the time within 1400~s from the trigger time (bottom panel, LAT data only)~\cite{080916c}.
{\it (Right)} GRB~090510 light curves from {\it Swift} and {\it Fermi} observations, given as energy flux densities averaged in the observed energy bands: BAT
(15-–350~keV), XRT (0.2-–10~keV), UVOT (renormalized to white), LAT (100~MeV–-4~GeV).
The prompt emission is shown for comparison: GBM (8~keV–-1~MeV, circles), LAT (100~MeV–-4~GeV, empty squares)~\cite{090510_afterglow}.
} 
\label{Fig:ExtEmission}
\end{figure}

In addition to beginning later than the low-energy emission, the high-energy emission in bright {\it Fermi}-detected 
bursts also appears to persist longer than the low-energy emission. The signature of this extended emission is
smooth and at a low flux level that would be difficult to detect in a background-limited instrument such as GBM, so that its
non-detection at low energies is not conclusive.
Certainly, as can be seen in the extended light curve
for GRB~080916C, left panel of Figure~\ref{Fig:ExtEmission}, the LAT emission (closed squares) persists for longer than is
obvious from the light curve in Figure~\ref{Fig:LC}, and its power-law decay over time (until it leaves the LAT field-of-view
at 23 minutes post-trigger) is more like an afterglow decay than the impulsive, pulsed structures that one associates
with the prompt emission in Figure~\ref{Fig:LC}.   
This association with afterglow emission is reinforced by the joint {\it Fermi} and {\it Swift} observations of GRB~090510, 
shown in the right panel of Figure~\ref{Fig:ExtEmission},  with a slight overlap in time between the latest LAT detection and
the earliest X-ray and optical data points.  The authors in~\cite{090510_afterglow} explore the relationship between
the extended LAT emission and the lower-energy afterglow emission from
GRB~090510, which is the only GRB observed jointly by the {\it Fermi} LAT and {\it Swift}.
The most extended emission seen in the LAT is for GRB~090328 where a smooth decay is seen up to $\sim$8~ks after the end of the impulsive prompt emission detected by GBM. This is the behavior most similar to that seen by EGRET for GRB~940217, but the occultation of the burst position to CGRO for the period between the prompt BATSE and EGRET episode and the high-energy photons detected by EGRET $\sim$75 minutes later prevent us from comparing the lightcurve of GRB~940217 to the smooth and continuous high-energy afterglow-like emission detected by the LAT for several GRBs.
\\
 
\begin{figure}[t!]
\includegraphics[width=0.45\linewidth]{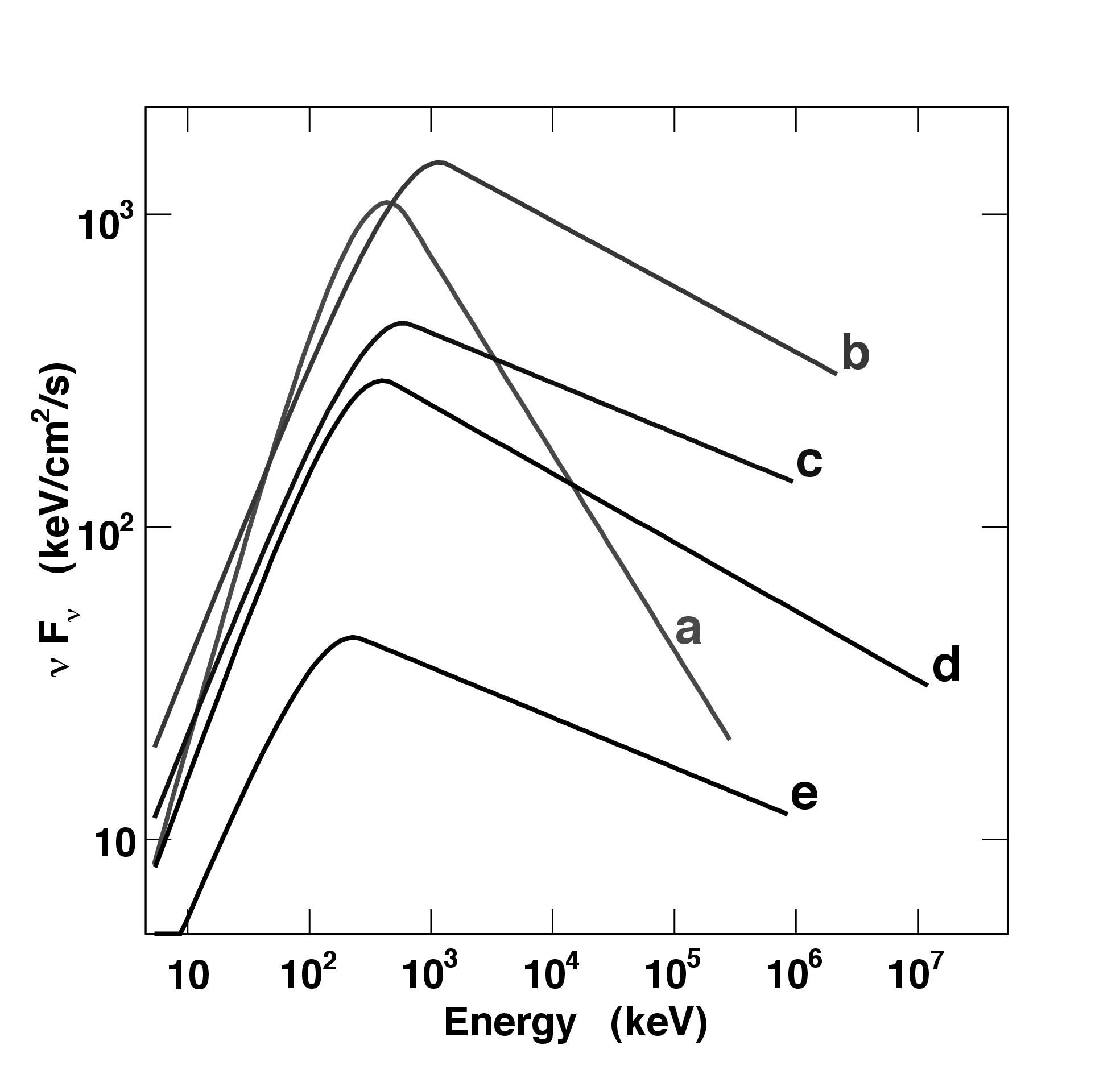}
\includegraphics[width=0.55\linewidth]{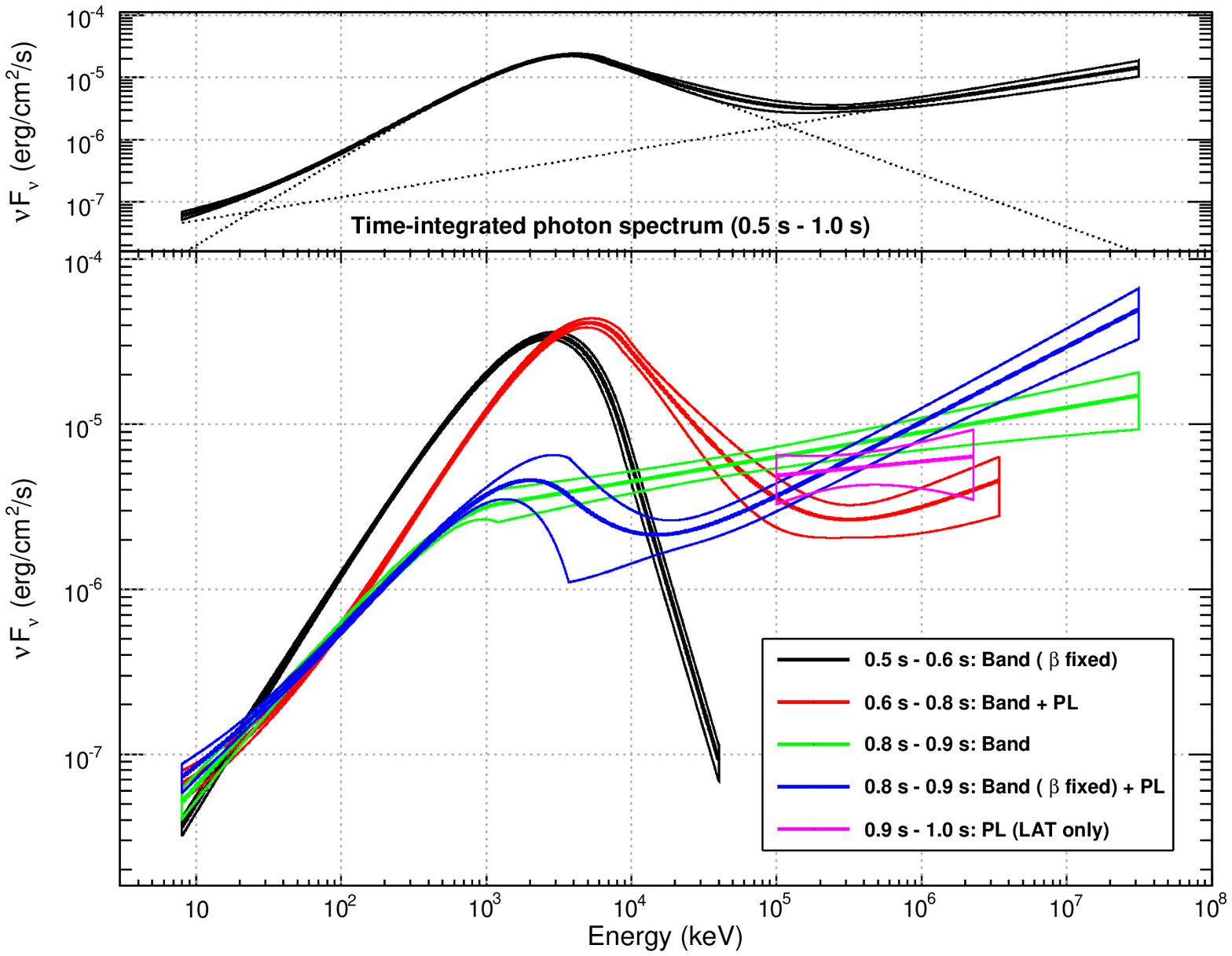}
\caption{
The model spectra of GRB~080916C (left)~\cite{080916c} and GRB~090510 (right)~\cite{090510_prompt} in $\nu F_\nu$ units, in which a flat spectrum would indicate equal energy per decade of photon energy.
The curves end at the energy of the highest-energy photon observed in each time interval.
\label{Fig:SpecEvol}
}
\end{figure}

The four bright LAT GRB detections offer the opportunity to explore
the energy spectrum of GRBs with unprecedented sensitivity over seven decades of energy,
revealing two broad types of behaviour.  In GRB~080916C the LAT energy spectrum extends from 8~keV to 13~GeV with no
obvious deviation from a single power law above $E_{peak}$ of the Band function~\cite{band93} in 
spectral fits from five time intervals. This simple functional fit, shown in Figure~\ref{Fig:SpecEvol} (left),
suggests a single physical mechanism dominates over a very broad energy range.  In the three other bright
LAT bursts, however,  an extra component is required in addition to the
Band function in order to account for the high-energy emission.  The time-integrated spectrum of
GRB~090510, displayed in the right panel of Figure~\ref{Fig:SpecEvol} (top), 
shows that the additional component can be fit by a power-law,  with the unexpected
result from {\it Fermi} that this component dominates not only
at high energies, above 100~MeV, but also low energies below 50~keV. 
Looking at the spectrum in slices of time (lower panel), it can be seen that the
Band function and power-law component are not both
required in each time interval. This raises the possibility that
the two components are not completely correlated, but unlike the
temporally separated low and high-energy components seen in the EGRET GRB~941017,
it is difficult to draw conclusions given the statistics of the time-resolved spectral fits.\\

With more bright LAT detections, these characteristics will provide a detailed picture of
the broad-band spectral and temporal properties of GRBs. The most surprising aspect so far is the similarity
between short and long GRBs in the behaviour of their high-energy emission.

\section{Physical Implications}
\label{Sec:Implications}
\subsection{Constraints from gamma-ray opacities}
\label{SubSec:GammaMin}
The variability and brightness of GRBs at high energies provide indirect but strong evidence for relativistic outflows as the sites of the observed prompt emission. 
For a source at rest, the short variability time-scales $t_v$ observed in the prompt light curves gives a limit on the size of the emitting zone $R<ct_{v}$, based on a simple causality argument.
Combined with the large luminosities $L\sim10^{50-53}$erg\,s$^{-1}$ inferred by assuming isotropic emission, this compactness is sufficient for photons of high energy $E$ to annihilate in pairs with dense fields of
softer photons with energies $\epsilon=E/m_ec^2\sim$1.
The huge implied optical depth $\displaystyle\tau_{\gamma\gamma}>10^{13}\left(\frac{L_{1/\epsilon}}{10^{51}\;\mathrm{erg\,s^{-1}}}\right)\left(\frac{t_v}{10\;\mathrm{ms}}\right)^{-1}$ would produce a thermal spectrum, in contradiction with the broad-band non-thermal power-law spectra observed up to high energies.
This well known ``compactness problem'' can be solved by considering a source moving at relativistic speed towards the observer\footnote{See the Supporting Online Material in~\cite{080916c} for a detailed computation.}.
In this case, the opacity is reduced by a factor $\displaystyle\Gamma^{2(1-\beta)}$, and can be less than unity for a typical slope $\beta\simeq-$2.3 of the high-energy power-law spectrum combined with a minimum value of the bulk Lorentz factor of the outflow $\Gamma>\Gamma_\mathrm{min}$$\sim$100 (increasing with $E$, $1/t_v$, the redshift and the source intensity).

\begin{figure}[t!]
\includegraphics[width=.55\linewidth]{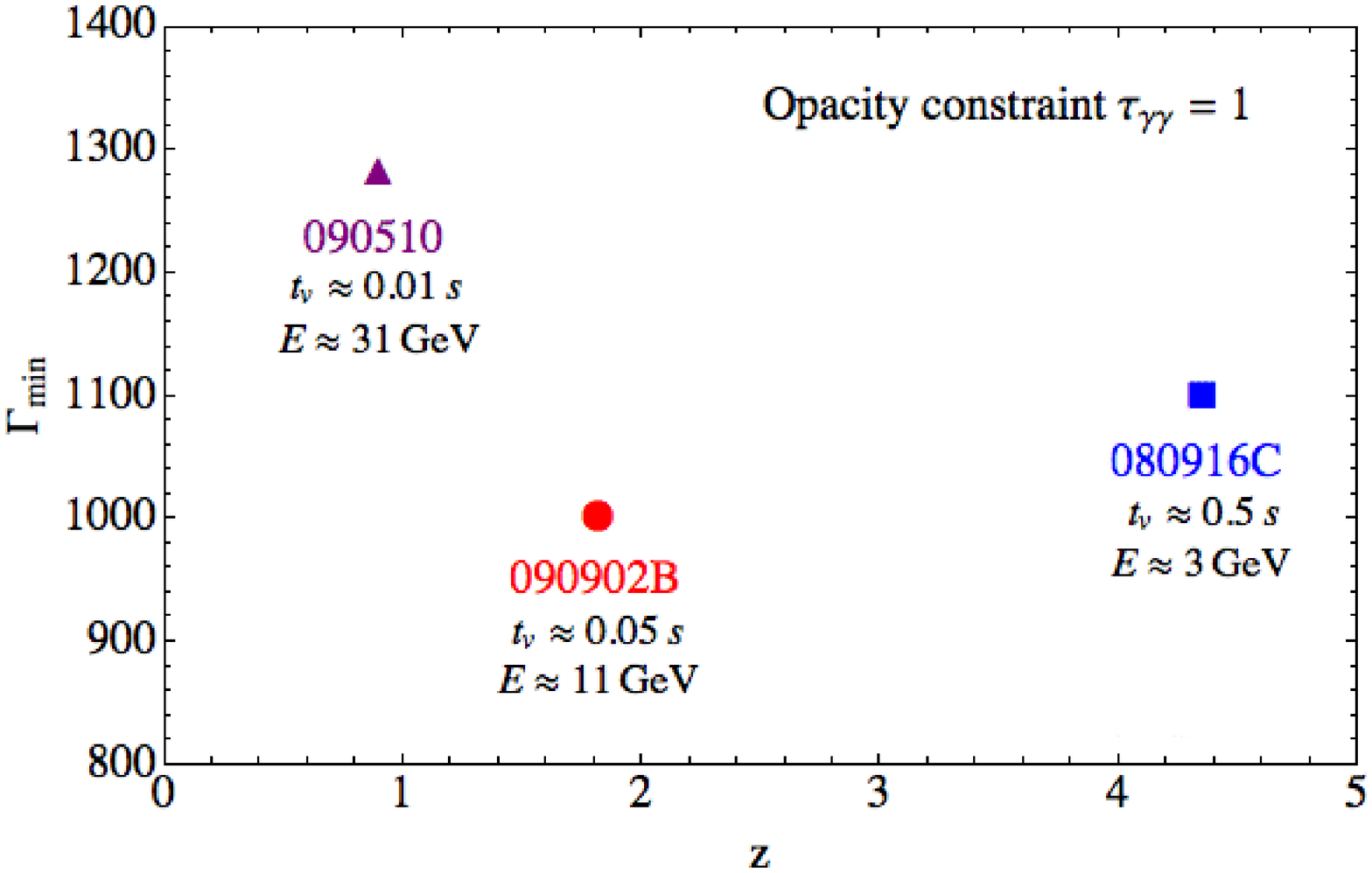}
\includegraphics[width=.45\linewidth]{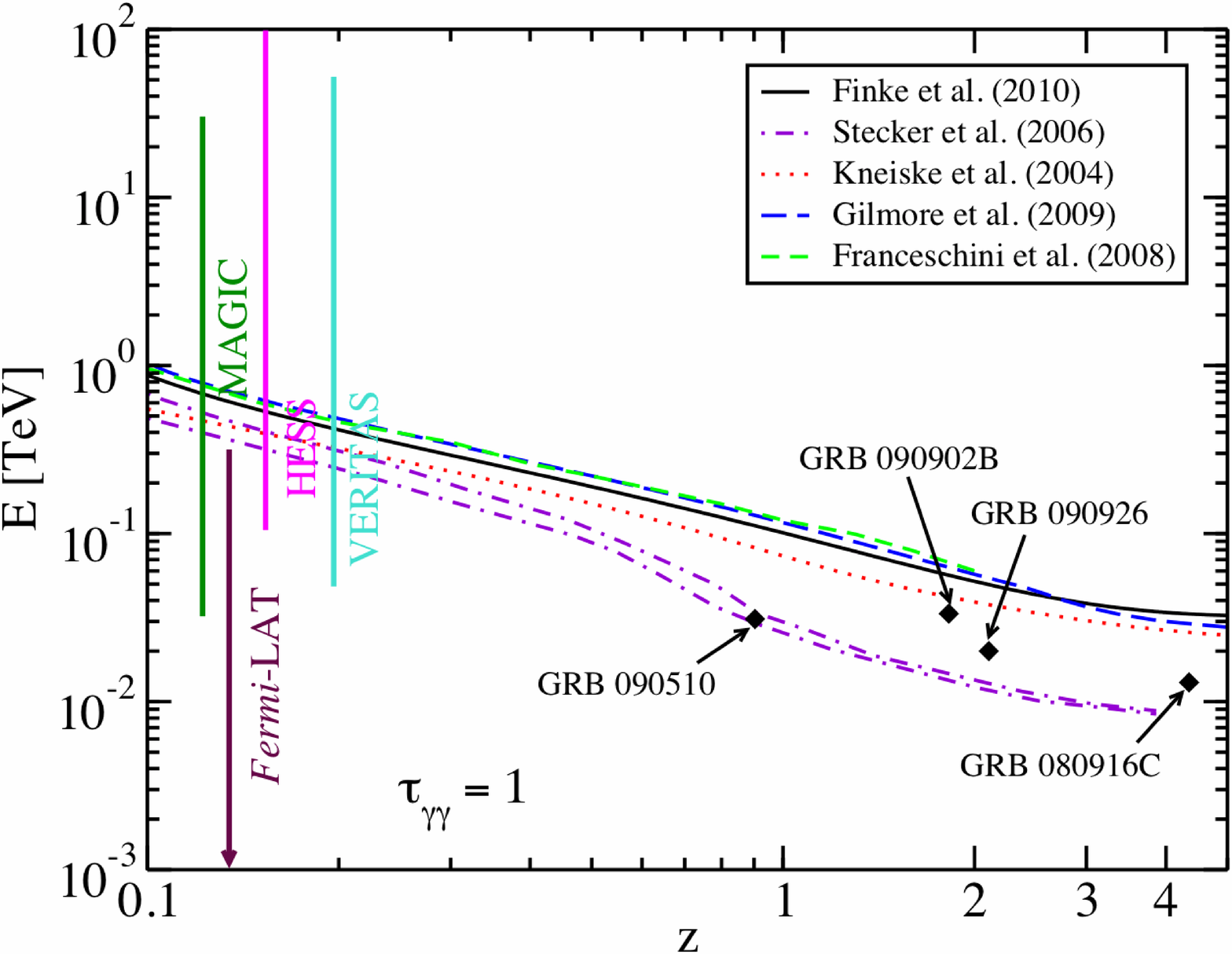}
\caption{
{\it (Left)} Lower limit $\Gamma_\mathrm{min}$ on the bulk Lorentz factor of the jet from LAT observations of the two long GRB~080916C and GRB~090902B, and the short GRB~090510, as a function of their redshift~\cite{080916c,090902b,090510_prompt}.
The limits are obtained by setting the opacity to pair production to unity, using a maximum photon energy and a variability time-scale as indicated for each case.
{\it (Right)} Constraints on the EBL from LAT detections of GRB photons above 10~GeV~\cite{EBL,finke}.
} 
\label{Fig:Opacities}
\end{figure}

Since the launch of {\it Fermi}, the LAT has detected photons with energies $E>10$~GeV from several very bright bursts.
These high-energy photons
 were used to set limits $\Gamma_\mathrm{min}\simeq1100$, and revealed that both long and short GRBs have high outflow Lorentz factors (Figure~\ref{Fig:Opacities}, left), a key result for GRB modelling.
These measurements are more robust than in the past since they do not assume that the spectrum extends beyond the highest-energy detected photon.
Some assumptions in these studies are, however, worthy of investigation and possible revision in the near future.
First, the identification of the soft photon spectrum is a delicate task, and ideally requires spectrosocopy with good statistics over the considered time-scale $t_v$.
In pratice, $t_v$ is chosen as the fastest variability observed in the GBM light curve (e.g., 50~ms for GRB~090902B, while some variability is observed in the LAT down to $\sim$90~ms~\cite{090902b}), and the target photon spectrum is derived over slightly larger durations. 
Moreover, the target photon field is considered uniform, isotropic and time-independent in this simple one-zone steady-state model.
More realistic computations~\cite{granot08}, which account for geometrical and dynamical effects, can lead to smaller opacities and thus to smaller lower limits (by a factor 2--3) on $\Gamma$.\\

In addition to intrinsic opacity, high-energy gamma rays can be absorbed by the Extragalatic Background Light (EBL) when travelling from the emitting region to the observer.
The EBL is a cosmic diffuse radiation field resulting from the emission of the first stars and its subsequent reprocessing by dust in the interstellar medium.
Knowledge of the evolution of the EBL's density with redshift is of great importance to the study of galaxy evolution and star formation in the early phases of the Universe.
Only photons with energies above $\sim$10~GeV suffer from pair creation on the EBL, and they can be used to probe the EBL as a function of redshift in the optical-UV range~\cite{EBL}.
The highest energy photons detected by the LAT from bright and distant GRBs provide lower limits on the opacity (here integrated along the line of sight), and thus useful constraints on the intensity of the EBL.
Figure~\ref{Fig:Opacities} (right) shows the iso-contours for an opacity equal to unity in the energy-redshift plane for several EBL models available in the literature.
As can be seen, this handful of photons (in particular the 33~GeV photon from GRB~090902B, at $z$=1.82) puts
 severe constraints on the models predicting the largest absorption, while most of the models remain optically thin to the highest-energy photons seen by the LAT over the observed GRB redshift range.
The phase space covered at $\sim$TeV energies by ground-based \v Cerenkov experiments is also displayed on the figure.
These telescopes are sensitive to closer extragalactic sources (essentially blazars), and probe the infra-red part of the EBL.

\subsection{Constraints on Lorentz Invariance Violation}

Bright transient events occurring at cosmological distances were
proposed in the late 1990's as powerful tools to study the quantum-gravitational nature of space-time and to test the existence of Lorentz Invariance Violation (LIV) as a consequence of some Quantum Gravity theories~\cite{gac98}.
In these theoretical frameworks, the speed of light in vacuum $v$ is no longer a constant and can effectively depend on the photon energy $E$.
The natural energy scale $E_\mathrm{LIV}$ at which this effect becomes dominant is generally considered of the order of the Planck energy scale\footnote{This constant can be obtained from a dimensional analysis based on the fundamental constants in quantum mechanics, special relativity and Newtonian gravity theories.}, $\displaystyle E_\mathrm{P}=\sqrt{\frac{\hbar c^5}{G}}\simeq1.22\times10^{19}$~GeV.
This energy is much larger than the energies involved in GRB experiments, thus for practical studies one considers a Taylor expansion of the photon dispersion relation
$\displaystyle p^2 c^2=E^2\:\Big[1+\xi\frac{E}{E_\mathrm{LIV}}+\mathcal{O}\left(\frac{E^2}{E_\mathrm{LIV}^2}\right)\Big]$,
which corresponds to the photon velocity $\displaystyle v=\frac{\partial E}{\partial p}\simeq c\:\left(1-\xi\frac{E}{E_\mathrm{LIV}}\right)$.
The experimental test of this fundamental physics law (Einstein's special relativity) consists of a search for dispersion effects in the arrival times of photons emitted together by the distant source.
The difference in arrival times of two photons with observed energy difference $\Delta E$ (appendix in~\cite{bolmont08}; see~\cite{jacob08} for the next terms of the series) is:
$\displaystyle \Delta t = \frac{1}{H_0} {\frac{\Delta E}{\mathrm{E}_\mathrm{LIV}}} \int_{0}^{z} \frac{(1 + z')\,dz'}{\sqrt{\Omega_\Lambda + \Omega_\mathrm{M} (1+z')^3}}$,
adopting a standard cosmology with flat expanding Universe and a cosmological constant ($\Omega_\mathrm{M}$$=$0.3, $\Omega_\Lambda$$=$0.7 and $H_0$$=$71$\:\mathrm{km}\:\mathrm{s}^{-1}\:\mathrm{Mpc}^{-1}$).

This time-of-flight technique was successfully applied to GRB~080916C, where a $\sim$13~GeV photon was detected by the LAT 16.5~s after the GBM trigger (Figure~\ref{Fig:LC}, left)~\cite{080916c}.
Assuming that the photon was emitted after the GRB trigger, this delay was taken as an upper limit on the dispersion that could be attributed to a LIV effect, and converted into a conservative lower limit $E_\mathrm{LIV}\gtrsim0.1\;E_\mathrm{P}$ for the sub-luminal case ($\xi=1$).
When applied to the $\sim$31~GeV photon detected 0.83~s after the GBM trigger from the short GRB~090510 (Figure~\ref{Fig:LC}, right), the same analysis yielded $E_\mathrm{LIV}>1.19\;E_\mathrm{P}$, and even stronger constraints
are possible if one adopts less conservative yet highly reasonable assumptions regarding the onset of the low-energy emission~\cite{090510_LIV}.
In addition, a similar and very robust limit $E_\mathrm{LIV}>1.22\;E_\mathrm{P}$ was obtained from the lack of smearing of the narrow spikes observed in the high-energy light curve of GRB~090510, 
a limit that is thus valid for both sub- and super-luminal cases ($\xi=\pm1$).
As a result, the LAT observations of $>$10~GeV photons from bright GRBs provide the best lower limits on $E_\mathrm{LIV}$ and strongly disfavour theoretical models which predict a linear variation of the speed of light with photon energy.

\section{Perspectives}
\label{Sec:Perspectives}

\subsection{Open questions}

The observations summarized in Section~\ref{Sec:Observations} pose interesting problems in the context of the simplest standard scenario, where the GRB prompt emission is produced by the acceleration and emission of high-energy particles at internal shocks in a relativistic outflow, and the afterglow emission results from its deceleration by the circumburst medium.
Substantial theoretical work has attempted either to improve the internal/external shock scenario of the fireball model, or to revise some of its aspects for a better match with the GRB temporal and spectral properties revealed by {\it Fermi}.

Leptonic models (e.g., electron synchroton emission or jitter radiation~\cite{medvedev00} at $\lesssim$MeV energies and inverse Compton or self-Compton processes at $\gtrsim100$~MeV energies) naturally predict the correlated variability at low- and high-energies observed in the {\it Fermi} light curves (Figure~\ref{Fig:LC}).
However, this first class of models needs fine tuning to produce a delayed onset of the highest energies which is longer than the spike widths, each pulse in the light curve marking a different shell collision and shock.
Such models also have difficulties producing the power-law excess below $\sim$50~keV, 
and to attribute physical meaning in terms of emission processes (i.e. synchrotron or inverse Compton emission) to the photon index value of the Band spectrum at low energy and of the spectrum in the LAT energy range.
However, theoretical extensions which include additional processes such as the photospheric component are promising and may provide a better agreement with these observed properties~\cite{ryde10,toma10,guiriec11}.

In hadronic models~\cite{asano09,razzaque09}, which investigate GRBs as possible sources of the Ultra-High Energy Cosmic Rays\footnote{See Chapter~3 of this report.}, the late onset of the $\gtrsim$100~MeV emission could result from the time needed to accelerate protons and ions and to develop cascades.
While subsequent proton synchrotron radiation requires large magnetic fields, synchrotron emission from secondary $e^+e^-$ pairs produced via photo-hadron interactions is possible and could explain the power-law excess below $\sim$50~keV.
On the other hand, these models do not naturally predict the aforementionned correlated variability, and they also require substantially more energy (1-3 orders of magnitude) injected in the fireball than observed.
However, this constraint strongly depends on the exact value of the bulk Lorentz factor of the outflow $\Gamma$, and could be accommodated with lower values of this key parameter (see Section~\ref{SubSec:GammaMin}).\\

Joint observations of GRBs with the GBM and the LAT shed new light on their possible emission mechanisms which, on average, radiate $\approx$20\% of their energy above $\sim$100~MeV.
The answers to the open questions outlined above, however,
need observations of more and brighter GRBs with both instruments in the near future, accompanied by more detailed
time-resolved spectroscopy, in order to pinpoint which high-energy processes dominate throughout the GRB.
In particular, the connection between the hard additional power-law component seen by the LAT in the prompt emission and the long-lived GeV emission observed up to several kilo-seconds, is of great importance
in understanding the transition from the internal shock phase to the early and late afterglow phases.

\subsection{Synergy with ground-based \v Cerenkov telescopes}
\begin{figure}[t!]
\includegraphics[width=\linewidth]{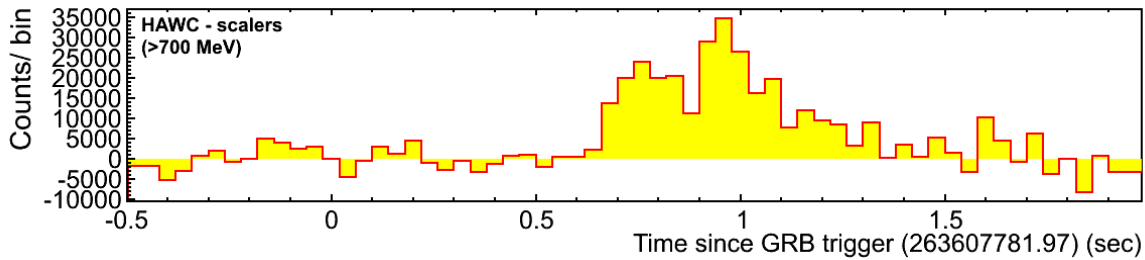}
\caption{
HAWC simulation of the count rate light curve for GRB~090510. The signal was summed over the 900 PMT scalers in the 300 water tanks of the ground-based array, and the average background has been subtracted.
The burst was placed at a small zenith angle, and its spectrum was convoluted by the EBL model from~\cite{gilmore09}, which attenuates the observed flux above $\sim$120~GeV (for a redshift $z$$=$0.9).
}
\label{Fig:VHE}
\end{figure}

The detection by the LAT of only a few percent of the GBM-triggered GRBs occurring in its field-of-view can be explained if one looks at the bursts detected by the LAT in the wider context of the GBM GRB population.
For both short and long GRBs, the LAT detects the events that are most fluent (for long GRBs) and have the highest peak flux (for short GRBs)
at lower energies, and have spectra that might reasonably extend to higher energies.
Although several LAT-detected GRBs require extra spectral components beyond the Band function detected at lower energies, they are still among the brightest, hardest bursts detected by GBM.
The relatively low detection rate by the LAT might then be set by the sensitivity of the LAT rather than
by an instrinsically small population of GRBs producing high-energy emission --
either because of internal opacity or because of some other mechanism resulting in a paucity of high-energy photons.\\

Although the chances of placing a more sensitive GeV telescope in orbit in the near future
are slim, the prospects for very high-energy (VHE) GRB astronomy on the ground are excellent.  
VHE observations of GRBs can be made both with the Imaging Atmospheric \v Cerenkov Telescopes (IACTs),
which have high sensitivity and low energy threshold (about 100--200~GeV for the current generation) 
at the cost of low duty cycle and narrow field-of-view,
and the less sensitive wide-angle water \v Cerenkov detectors which can operate with nearly 100\% duty cycle.
The story of IACT observations of GRBs has so far been one of upper limits, probably because of the finite response time
of the narrow field telescopes to GRB trigger notifications from space instruments, the poor localization
capabilities of the most prolific bright GRB detectors (BATSE and GBM), the faintness of most of the GRBs seen by
the instrument with both numerous triggers and good localization capabilities ({\it Swift}), and the energy threshold and sensitivity of VHE searches to date. 
Upper limits to late-time emission have been published for dozens of GRBs followed up by MAGIC~\cite{magic}, VERITAS~\cite{veritas},
and HESS~\cite{hess}, with typical response times of tens of seconds, 1-2 minutes and hours, 
respectively. None of the observed bursts was detected by the LAT so that the lack of detected late-time VHE emission
cannot be placed in context with the late-time GeV emission seen in the LAT. The overlap in energy coverage between
the LAT and these IACTs, particularly MAGIC with an energy threshold of 25~GeV for its special
GRB observation mode, combined with the priority given to GRB follow-up observations by the VHE IACT community
given the detection of late-time emission in the LAT at GeV energies,  offers the possibility
of a joint LAT-IACT detection with the current generation of IACTs over the next few years.  Owing to the need for IACTs to slew in response
to a trigger from another instrument, onboard LAT triggers are especially desirable,
and the response time of VERITAS and MAGIC is well-matched to the time-scale
of the extended emission seen in LAT-detected GRBs.\\

The bright GRB~090902B was particularly inspiring to the VHE community given the detection of the highest energy photon ever seen from a GRB,
33~GeV, 80 seconds after the GBM trigger time.  
The next generation IACT experiment, the more
sensitive \v Cerenkov Telescope Array (CTA;~\cite{cta}) will, under the current schedule, operate during the lifetime of
{\it Fermi}, and offers the best hope for VHE observations of GRBs that capitalize on the successes of the {\it Fermi} LAT
in characterizing the extended emission of GRBs.
It is essential to design CTA with the lowest possible energy threshold, so that absorption by the EBL does not prevent us from sampling GRBs over a range of redshifts.
In addition to providing insights regarding emission mechanisms in these extreme objects, the GRB spectra seen by CTA could help us to discriminate between EBL models at distances farther than possible with measurements of blazars.

For VHE observations of the prompt emission in GRBs, water \v Cerenkov detectors such as the MILAGRO experiment,
which operated until 2008, offer the best chance of success because the brightness of the prompt phase relative to
the late-time emission (Figure~\ref{Fig:LC}, left) may overcome the lack of sensitivity relative to the IACTs and a
serendipitous overlap is quite likely given the field-of-view and duty cycle of these experiments.  A statistically unlikely
clustering was detected by the MILAGRITO
prototype for MILAGRO in coincidence with BATSE GRB~970417A~\cite{milagrito},
but no counterparts were seen to GRBs with the more sensitive MILAGRO~\cite{milagro}, making this
association questionable or suggesting such bright GRB VHE emission is rare.  A next generation water \v Cerenkov detector, HAWC~\cite{hawc}, 15 times
more sensitive than MILAGRO, is currently under construction in Mexico and will be operational during the {\it Fermi} and SVOM eras. 
Post-{\it Fermi}, a HAWC trigger may provide the best real-time notification for IACTs that a GRB has occurred which is likely to have
an extended VHE signal (Figure~\ref{Fig:VHE}). Together, HAWC, CTA and SVOM will detect, localize and characterize GRBs over a huge energy range,
building on the success of the {\it Fermi} mission.





\end{document}